\documentclass[prb,
twocolumn,
superscriptaddress,showpacs,amsmath,amssymb]{revtex4}

\usepackage{hyperref}
\usepackage{amsfonts}
\usepackage{bm}
\usepackage{verbatim}
\usepackage{color}
\usepackage{graphicx}
\usepackage[applemac]{inputenc}

\newcommand{\tr}{\textrm{tr}}
\newcommand{\Tr}{\textrm{Tr}}

\newcommand{\im}{i}

\newcommand{\vek}[1]{\mathbf{#1}}
\newcommand{\doo}{\partial}

\begin{document}

\title{Field theory of higher-order topological crystalline response, generalized global symmetries and elasticity tetrads}

\author{Jaakko Nissinen}
\email{jaakko.nissinen@aalto.fi}
\affiliation{Low Temperature Laboratory, Department of Applied Physics, Aalto University,  P.O. Box 15100, FI-00076 Aalto, Finland}

\date{\today}

\begin{abstract}
We discuss the higher-order topological field theory and response of topological crystalline insulators with no other symmetries. We show how the topology and geometry of the system is organised in terms of the elasticity tetrads which are ground state degrees of freedom labelling lattice topological charges, higher-form conservation laws and responses on sub-dimensional manifolds of the bulk system. In a crystalline insulator, they classify higher-order global symmetries in a transparent fashion. This coincides with the dimensional hierarchy of topological terms, the multipole expansion, and anomaly inflow, related to a mixed number of elasticity tetrads and electromagnetic gauge fields. In the continuum limit of the elasticity tetrads, the semi-classical expansion can be used to derive the higher-order or embedded topological responses to global U(1) symmetries, such as electromagnetic gauge fields with explicit formulas for the higher-order quasi-topological invariants in terms of the elasticity tetrads and Green's functions. The topological responses and readily generalized in parameter space to allow for e.g. multipole pumping. Our simple results further bridge the recently appreciated connections between topological field theory, higher form symmetries and gauge fields, fractonic excitations and topological defects with restricted mobility elasticity in crystalline insulators.
\end{abstract}

\maketitle

\section{Introduction}

Higher order topological insulators generalize symmetry protected topological phases in spatial $d$-dimensions and are protected (usually) by spatial symmetries of crystalline origin. They harbor protected edge excitations of higher codimensionality: on the $d-1$-dimensional boundary, $d-2$- or lower-dimensional modes robust to symmetry preserving perturbations occur. In particular, multipole insulators with localized boundary moments, i.e. localized dipoles, quadrupoles etc. have been considered \cite{Benalcazar17b, Benalcazar17, SchindlerEtAl18}. 

Their close relation to subsystem \cite{HughesEtAl19} and higher-order gauge symmetries in field theories has been recently noted and explored in various directions \cite{YouEtAl19, SeibergShao20, GorantlaEtAl20}. This connection was appreciated by another recent development, where the notion of higher-form global and gauge symmetries has been considered as a convenient organizing principle for the phase structure and the spectrum of allowed operators in (gauge) field theories \cite{GaiottoEtAl15}. In fact, the essential physics following from this organizing principle were implicitly contained in the early, seminal Refs. \cite{Polyakov75}. More generally one expects, also in gauge theories, that global symmetries are more useful than local ones in analyzing the phase structure, since the latter are really redundancies in the degrees of freedom. The higher-form generalized global symmetries couple to line, surface, hypersurface and volume operators instead of the conventional global symmetries that couple to point operators. In such theories, conserved charges are, respectively, measured similarly on spatial hypersurfaces of suitable codimension. Systems with such symmetries in topological phases with long-range entanglement have been shown to have excitations with restricted mobility. Such excitations are now referred to as fractons, see e.g.\cite{FractonReview, Pretko17, PremEtAl18, PretkoEtAl20, GromovEtAl20}. In another recent development, elasticity, or more precisely its dual formulations, have been recast as theory of topological defects with fractonic character and higher-form conservation laws. Indeed the work on dislocation mediated melting is a (lattice) version of the confinement problem with a dual Higgs phase: the stress superconductor with confinement of dual, stress magnetic fields \cite{Kleinert, BeekmanEtAl17, BeekmanEtAl17b, PretkoRadzihovsky18}. 

On the other hand, the geometric response of topological crystalline systems in terms of finite translational lattice gauge fields, have been considered recently \cite{NissinenVolovik2019, Vishwanath2019}. Given the lattice, these fields are of tautological nature themselves but can lead to non-trivial symmetries in transport, and appeared also in \cite{PoovuttikulGrozdanov18} in the context of magnetohydrodynamics. Namely, the weakly protected 3+1d quantum Hall response, topological electric polarization and the associated anomaly inflow structures were considered \cite{NissinenVolovik2019, Vishwanath2019, NissinenHeikkilaVolovik20}. The anomalies are, respectively, the mixed axial-gravitational and Luttinger-type anomaly of the boundary theory. In both cases, lattice dislocations carry topological charges. In three spatial dimensions, these two responses are dual to each other. The former singling out one-dimensional lattice directions (1-cycles) $P_a$ and the second one two-dimensional (2-cycles) with normal $N_a$. Such a formulation is reminiscent of the higher-form symmetries and currents where, in the electromagnetic response, all fields couple explicitly to the electromagnetic fields with the ordinary gauge and global charge conservation symmetry. 

Motivated by this analogy, here we will consider the other classes of U(1) responses related to subsystem and higher-order symmetries in crystalline systems. We will look at the possible forms of insulators with additional higher-form crystalline global conservation laws. The expectations is that the higher-form responses can be derived from the semi-classical expansion in momentum space, much like the momentum space invariants of 0-form topological insulators \cite{QiHughesZhang08}. 

Concerning the crystalline lattice, we will phrase our results in the continuum formulation of elasticity and cohomology, expecting that the end results with topological character do not qualitatively change upon the incorporation of the discrete lattice theory, space or point group symmetries and discrete (co)homology. For simplicity, we focus exclusively on spatial lattices although our results can straightforwardly extended to driven time-periodic (Floquet) systems.

The rest of this paper is organized as follows. In Sec. \ref{sec:higher-order}, we review higher-order symmetries in U(1) Maxwell gauge theory which is sufficient for our purposes and connects with the responses to EM fields in the presence of higher-order global crystalline symmetries. Then we discuss the main application, (continuum) lattice field theory with higher-form crystalline global symmetries. The connection to the EM response in higher-order topological crystalline insulators is made in Sec. \ref{sec:generalized_response}. The associated momentum space invariant are discussed in \ref{sec:momentum_space}. We end with Conclusions and Outlook.

\emph{Notations and conventions.} For ease of notation, throughout we employ differential form notation. For example, the electromagnetic 2-form $F= dA = \frac{1}{2} F_{\mu\nu} dx^{\mu} \wedge dx^{\nu}$ with $A = A_\mu dx^{\mu}$ the 1-form EM gauge potential. $\star$ is the spacetime Hodge dual, mapping $n$-forms to $d+1-n$-forms. The $(d+1)$-dimensional metric is set to $g_{\mu\nu} = \eta_{\mu\nu} = \textrm{diag}(-1,1,1,\dots,1)$, leading to $\star^2 = -(-1)^{n(d-n)}$ for $n$-forms. Although the construction is general, we mostly specialize to 3+1 dimensions and work in units where $e^2=\hbar=1$. The electric and magnetic fields are denoted as $\vek{\mathcal{E}}_i = F_{0i}$ and $\vek{\mathcal{B}}_i = \epsilon^{ijk}F_{jk}$ and the lattice one-forms $E^a = E^a_{\mu} dx^{\mu}$ are the (continuum) elasticity tetrads. 

\section{Review of higher form symmetries}\label{sec:higher-order}

We start by reviewing higher-form symmetries \cite{GaiottoEtAl15, GaiottoEtAl17, PoovuttikulGrozdanov18}, focusing to a relevant example for our applications: of ordinary U(1) Maxwell gauge theory in 3+1d. In the following we shall eventually utilize the global higher form symmetries in insulating crystalline systems that are inherited from the EM charge conservation. 

In $d+1$-dimensions, $p$-form global symmetries are an extension of the usual global symmetries with point like, i.e. 0-form, charges. The existence of a $p$-form symmetry leads to the existence of conserved $p+1$-form currents. The charged objects are $p$-dimensional with charges measured on $d-p$-dimensional surfaces. Similar to ordinary global symmetries, $p$-form can be spontaneously broken in $d-p>2$ dimensions, can be anomalous on their own or when gauged in quantum theory \cite{GaiottoEtAl15}. 

In 0-form U(1) theory with field $\phi \to \phi + \lambda$, $q$-charged point-operators are transformed as $O_{q}(x) \to e^{\im q \lambda}O_{q}(x)$. In 1-form symmetric theory, $Q$-charged operators are Wilson loops transforming as $W_Q[C] \to e^{\im Q \int_C \Lambda} W_Q[C]$ under global gauge transformations $A\to A+\Lambda$, where $A, \Lambda$ are the elementary abelian 1-form charged field and shift parameter.

Let us consider (3+1)d U(1) $0$-form gauge theory as an example. The gauge action is
\begin{align}
S_{\rm EM}[A, j] &= \int \sqrt{g} d^4 x \frac{1}{4e^2} F_{\mu\nu}F^{\mu\nu} + A_{\mu} j^{\mu} \nonumber\\ 
&= \int F \wedge \star F + A\wedge \star j \label{eq:Maxwell}
\end{align}
where $F\equiv dA$ and $\star$ is the spacetime Hodge dual. The 1-form $j \equiv j_{\mu} dx^{\mu}$ represents \emph{electrically} charged matter. The equations of motion are
\begin{align}
d \star F = \star j, \quad dF = 0 \label{eq:EM_EOM}
\end{align}
where the latter is a topological Bianchi identity from $d^2A=0$, tantamount to 0-form gauge symmetry of $A \to A + d\lambda$. The corresponding current is conserved
\begin{align}
d\star j = \doo_{\mu}j^{\mu} = 0. 
\end{align}
This statement is equivalent to the (global and local) 0-form charge conservation
\begin{align}
\frac{dq}{dt} = \frac{d}{dt} \int_{\rm space} \star j  =0. \label{eq:charge_conservation}
\end{align}

The Wilson loops are $\langle W_A[C] \rangle = \int DA \exp{\left[ \im \oint_C A \right]} e^{-\im S[A]}$ and their asymptotic behaviour is diagnostic for the phase and charged spectrum. However, and not unrelated to this, since local gauge symmetries represent redundancies in the physical degrees of freedom, the properties of the theory are better understood from 1-form global symmetries. These are given by 2-form currents $J_{e} = F$ and $J_{m} = \star F$, or
\begin{align}
J_{e\mu\nu} = \frac{1}{2} F_{\mu\nu}, \quad J^{\mu\nu}_{m} = \frac{1}{4} \epsilon^{\mu\nu\lambda\rho} F_{\lambda\rho} 
\end{align}
Now from the equations of motion \eqref{eq:EM_EOM}, 
\begin{align}
d J_m = \star j, \quad d J_e = 0, \label{eq:EM_2-form}
\end{align}
leading to, over two-dimensional spatial surfaces $\Sigma^{(2)}$,
\begin{align}
Q^{(2)}_{m} &= \frac{1}{2\pi} \int_{\Sigma^{(2)}} \star J_m =  \frac{1}{2\pi} \int_{\Sigma^{(2)}} F = \mathbb{Z}\\
Q^{(2)}_{e} &= \frac{1}{2\pi} \int_{\Sigma^{(2)}} \star J_e = \frac{1}{2\pi} \int_{\Sigma^{(2)}} \star F \\
& = \frac{1}{2\pi} \int_{\Sigma^{(2)}} \vek{E}\cdot d\vek{S} = \int_{\Sigma^{(3)}} j^0. \label{eq:electric_flux2}
\end{align}
Physically $Q^{(2)}_m$ is just the magnetic flux, which is conserved due to the Bianchi identity (magnetic field lines are closed; absence of magnetic monopoles). On the other hand $Q^{(2)}_e$ is the electric flux, which is not conserved (electric field lines end on charges). On the last equality of \eqref{eq:electric_flux2}, we used $\doo \Sigma^{(3)} = \Sigma^{(2)}$ and \eqref{eq:EM_2-form}. It is important to consider compact and closed spatial manifolds for the charges, since they are well-defined without additional boundary conditions. For example, if $\Sigma^{(3)}$ is the total spatial volume, the charge is of course conserved by \eqref{eq:charge_conservation}. Similarly, for topologically non-trivial spaces, we can transform e.g. a line operator $\exp[\im \frac{1}{2\pi}\oint_{C} A] = \exp[\oint_{C} \frac{\lambda}{2\pi}] W_{A}[C]$ on a closed homology loop with a large gauge transformation, so that $Q^{(2)}_{m}$ is still conserved. 

We see that if there are no free, light charges $j=0$ in the vacuum, electric and magnetic fields become symmetric. The 2-form currents and conservation laws are $\star$-dual and $\star F = d\tilde{A}$ for a magnetic photon $\tilde{A}$. The usual Coulomb phase corresponds to the the phase with magnetic 2-form symmetry, whereas confinement adds the electric one (as relevant e.g. in \cite{Polyakov75} or non-abelian gauge theory). Likewise, proliferation of (dual) magnetic/electric vortex operators can break, or restore, the symmetry to, or from, a subgroup. On the other hand, the usual electromagnetic duality can be also extended to the case where we consider magnetic monopoles $\tilde{j}$ in addition to $j$.

We can now ask the following question: what phases of matter can lead to approximate higher-form symmetries related to the local U(1) gauge group of electromagnetism and how will their EM response look. Essentially, we look for the ways the light (in a suitable sense) electric charges can be excluded from the vacuum. Next we shall discuss this and higher-form global symmetries in crystalline insulators, where we have massive free charges at low enough energies below some mobility gap.

\section{Crystalline topology of elasticity tetrads: almost conserved higher order gauge symmetries} \label{sec:elasticity_tetrads}

\subsection{Lattice geometry in the continuum limit}

In $d$-dimensions, a lattice $L$ can be embedded to space as system of $d$ crystallographic coordinate planes of constant phase $X^a(x)$, i.e. $e^{\im X^a(x)}=e^{\im 2\pi n^a}$, $n^a \in \mathbb{Z}$ with $a=1,\dots, d$ \cite{AndreevKagan84}. In $d=3$, the intersections of the surfaces
\begin{equation}
X^1({\bf r},t)=2\pi n^1 \,\,, \,\,  X^2({\bf r},t)=2\pi n^2 \,\,, \,\, X^3({\bf r},t)=2\pi n^3 \,,
\label{points}
\end{equation}
then define the (possibly deformed) crystal lattice 
\begin{align}
L = \{ \mathbf{r} = {\bf R}(n_1,n_2,n_3) \vert \mathbf{r}\in \mathbb{R}^3, n^a \in \mathbb{Z}^3\}. \label{eq:lattice}
\end{align}
Instead of the periodic scalars $X^a$, the elasticity tetrads $E^{~a}_\mu(x) = \doo_{\mu} X^a$ represent the conventional hydrodynamic variables of elasticity theory \cite{DzyalVol1980, NissinenVolovik2019}. The depend slowly on the spacetime coordinates and encode the geometry and topology of the lattice embedded in spacetime. Note that this lattice has not necessarily anything to do with the underlying crystal lattice of the electronic insulator but instead is a ground state property of the coarse-grained state \cite{Oshikawa00}. The lattice $X^a\to X^a+2\pi$ symmetries are tantamount to the existence the translational gauge fields, the elasticity tetrads.

In $d=3$, the elasticity tetrads are gradients of the three U(1) phase fields $X^a$, $a=1,2,3$,
\begin{equation}
E^{~a}_\mu(x)= \partial_\mu X^a(x)\,
\label{reciprocal}
\end{equation}
and have units of crystal momentum. While the index $a$ is understood to be spatial, the $\mu =t,x,y,z$ for time dependent deformations, with velocities $V^a = \doo_t X^a$. Moreover, in order to couple to EM, we have to assume the continuity equation of charge conservation. For small deformations $X^a =  \frac{2\pi \delta^i_a}{\ell^a}(x^i + u^i)$, leading to $E^a_{\mu}= \frac{2\pi}{\ell^a}(\delta_\mu^a+\doo_\mu u^a)$. The currents from time-deformations are
\begin{align}
\doo_t \rho + \nabla\cdot j = \rho_{\rm u.c.}(-\doo_t \doo_i + \doo_i \doo_t) u^i = 0.
\end{align}
where ``u.c." denotes unit-cell averaged (charge) density. This conservation law holds in the crystal in the absence of interstials/vacancies, being equivalent to the dislocation glide constraint (no motion parallel to the Burgers vector). In the absence of dislocations, the $X^a(x)$ are globally well-defined, meaning that the tetrads $E^{~a}_\mu(x)$ are pure gauge and satisfy the general integrability condition: 
\begin{equation}
T^a = d E^a = \frac{1}{2}(\partial_\mu E^{~a}_\nu-\partial_\nu E^{~a}_\mu) d x^{\mu} \wedge d x^{\nu}=0.
\label{integrability}
\end{equation}
In the rest of the paper, we assume $V^a = 0$, focusing on the electronic couplings proportional to $\vek{\mathcal{E}}$ in a (dielectric) crystalline insulator with localized charges. Effects of non-zero velocity would also induce magnetic terms, proportional to $\vek{V} \times \vek{\mathcal{B}}$ and corresponding to terms $V^a \wedge E^b \wedge dA$ below. This extension has to be made bearing in mind that the dual form coupling to $p$-form currents and charges change their dimensionality when time-like components in $d+1$-dimensions are included.

The basis $E^a_{\mu}$ is then simply made non-degenerate by adding $E^0_\mu = \frac{1}{T}\delta^0_{t}$, where $T$ is time-periodicity and letting $T\to \infty$ at the end. In the simplest undeformed case, $X^a(\mathbf{r},t) = \mathbf{K}^{a} \cdot \mathbf{r}$, where $E_i^{(0)a} \equiv \mathbf{K}^a$ are the (primitive) reciprocal lattice vectors $\mathbf{K}^a$. In the general case, they depend on space (and time) but are still are quantized in terms of the lattice $L$ in Eq. \eqref{eq:lattice}.

By the inverse function theorem, we can define the inverse vectors,
\begin{align}
E^{~a}_{\mu}(x) E_{~a}^{\nu}(x) = \delta^{\mu}_{\nu},
\end{align}
and the lattice metric $G$ associated with these tetrads,
\begin{align}
G_{ab} = E_{\ a}^{\mu} E_{\ b}^{\nu} \eta_{\mu\nu}, \quad G^{ab} = E^{\ a}_{\mu} E^{\ b}_{\nu} \eta^{\mu\nu}, \label{eq:lattice_metric}
\end{align}
where $\eta$ is the metric associated to the background spacetime, say the spatial Euclidean (or the four-dimensional Minkowski metric). Note that $dn^2 = G_{ab} E^{a} E^{b}$ \emph{counts} distances in terms of the \emph{number of} lattice points of $L$ \cite{AndreevKagan84}. We assume a simple cubic or orthorombic lattice. Loosely speaking, the elasticity tetrads are the trivial gauge fields corresponding to the $U(1)^d$ translational symmetries along these directions.

\subsection{Topology}

Now we describe the differential topology of this lattice and embedding in detail. Upon making the identifications $\mathbb{T}^d \simeq \mathbb{R}^d / L$, we can compute the holonomies
\begin{align}
\langle C_b, E^a \rangle= \int_{C_b} E^a = 2\pi n^a\delta_{ab},\quad a,b=1,\dots,d 
\end{align}
where we fix the origin at $X^a = 0$ and there are $n^a = n^a(L)$ lattice points along the direction $C_a = \{X^b = \textrm{const.}\vert a\neq b \}$. The lattice $L$ in \eqref{eq:lattice} is then defined by the trivial holonomy sections $e^{\im \langle C_b(x), E^a \rangle} = 1$, where $C_b(x)$ is the open 1-cycle from the origin to the point $x$.

Similarly, the flat gauge $T^a = dE^a =0$ fields on the torus $\mathbb{T}^d$ take the form
\begin{align}
E^a = \mathbf{K}^a(x)\cdot d\mathbf{x} + \tilde{E}^a(x)
\end{align}
where $\mathbf{K}^a(X^a) = \frac{2\pi}{\ell^a}$ are the lattice directions of periodicity, the generators of the 1-cohomology group of $\mathbb{T}^d$. The smooth gauge part $d \tilde{E}^a(x) = 0$ is arbitrary and does not depend on the lattice topology, $\tilde{E}^a(x) = \doo_{\mu}\tilde{X}^a$, where $\tilde{X}^a$ is smoothly connected to identity. We conclude that the $E^a, C_a$ are simply $H^1(\mathbb{T}^3, \mathbb{Z})$ and $H_1(\mathbb{T}^3,\mathbb{Z})$ embedded via $X^a(x)$ in the spacetime and suitably normalized. 

In the presence of dislocations, $T^a \neq 0$, and the embeddings $X^a(x)$ are multivalued. Then we can define the embedding by also including the operators on 2-cycles $C_a \cup C_b $, 
\begin{align}
\int_{C_{b} \cup C_{c}} T^a = 2\pi B^a \label{eq:torsion_dislocation}
\end{align}
where $B^a$ is the density of Burgers vector. It can be shown that the lattice with dislocations is connected by a set of large gauge transformations to the defect-free lattice. The broken translational symmetry of the lattice can be restored when such vortex operators proliferate or condense. In this context we note that the rotational defects, i.e. disclinations have considerably higher energies. As such they only occur as bound dipoles in crystals in internal equilibrium, which allows us to set the spin-connection $\omega^{a}_{\ b}$ to zero in the coarse-grained continuum limit. 

\subsection{Higher form crystalline symmetries}

The connection of elasticity, dislocations and disclinations to higher form symmetries and gauge theories has recently attracted attention. In particular, the restricted motion of dislocations in terms of the glide constraint can be understood as fractonic excitations of higher form (dual) gauge theories. From this perspective, the dislocation glide constraint is protected by the 0-form U(1)-number conservation (in effectively neutral atomic crystal), leading to the higher form conservation of dipole moments in the higher-rank dual gauge theory.

Motivated by this and the bulk-boundary correspondence in topological insulators, we now wish to consider the localized (electronic) charges in a crystalline background and their relation to the global U(1) charge conservation. Such a consideration is motivated by the intuitive picture of weak crystalline topological insulators, both as subdimensional insulators and higher-form topological insulators. In the context of higher-order topological crystalline insulators, we already expect states with non-trivial higher order boundary responses and charges with restricted motion to be related to higher-form symmetries. In this case, we want to consider the effects of the crystalline symmetries in the insulator in combination with the local U(1) gauge symmetry of electromagnetism.

We imagine that the crystalline insulator is composed of unit cells with integer number of charges, say with a simple cubic or orthorombic symmetry. Since the system is an insulator, the (electronic) charges are localized in the unit cell below a mobility gap, even in the presence of external fields. The simplest occurrence of insulation is when the unit-cell (u.c.) (or ``voxel" \cite{Benalcazar17}) coarse-grained charge density $\rho_{\rm u.c.}$ is (locally) conserved
\begin{align}
\frac{d}{dt} \int_{\rm space} \rho_{\rm{u.c.}} = \int_{\rm space} \nabla\cdot \vek{j}_{\rm{u.c.}} = 0
\end{align}
the last equality over total volume is non-trivial with open boundary conditions only. In addition, and less constraining, it might be that \cite{GromovEtAl20}
\begin{align}
\frac{d}{dt} \int_{\rm surfaces} \rho_{s} &= 0, \nonumber\\
\frac{d}{dt} \int_{\rm lines} \rho_{\rm l} &= 0, \label{eq:lattice_higher_form}
\end{align}
etc. in higher dimensions. Here $\rho_s$ and $\rho_l$ are really charge densities per unit plaquettes and links on the lattice. Above we have specialized to three dimensions, although the conservation of hypersurface charges is an easy generalization. In terms of the elasticity tetrads, the conservation laws are written respectively as, $\det(E^a)\sim\rho_{\rm u.c.}$,
\begin{align}
d( E^1 \wedge E^{2} \wedge E^3) = 0 \quad \textbf{[0-form]} \label{eq:0-form}
\end{align}
and, for $a,b,c=1,2,3$ in three dimensions,
\begin{align}
d(\frac{1}{2}\epsilon_{abc} E^b \wedge E^c)=0, \quad &\textbf{[1-form]} \label{eq:1-form}\\
dE^a = 0, \quad &\textbf{[2-form]} \label{eq:2-form},
\end{align}
where we have indicated the relevant global $p$-form symmetry for clarity. These equations are of the form
\begin{align}
d(* J^{(p)}) =0 \label{eq:lattice_Hodge}
\end{align}
where now the duality $*$ is the Hodge star on elements of $\bigwedge^{k} E^a$, incorporating the lattice metric \eqref{eq:lattice_metric}. We note the dual nature of the currents
\begin{align}
*1 = E^1\wedge E^2\wedge E^3, \quad
*E^a = \frac{1}{2} \epsilon^{a}_{\ bc} E^b \wedge E^c. \label{eq:dual_pairs}
\end{align}
we see that the ''most trivial" conservation law in the crystalline insulator is actually the 3-form volume conservation law
\begin{align}
d1 = d(* E^1 \wedge E^2 \wedge E^3) \quad \textbf{[3-form]}.
\end{align}
This is reminiscent of the non-trivial 2-form \emph{electric} conservation law in U(1) Maxwell theory, essentially dictated by constrained dynamics.  In general $d$-spatial dimensions, such $p$-form symmetry current conservation laws would look like
\begin{align}
d (* J^{(p)}) = d(\frac{1}{n!}\epsilon_{a_1\dots a_d} E^{a_{p+1}} \wedge \dots \wedge E^{a_{d}}) = 0.
\end{align}
and we can write down a hierarchy of $p$-form currents and symmetries along specified lattice hypersurfaces. We will discuss below the responses corresponding to these conservations laws.

We note that the conserved elastic currents are tautological since the $E^a$ are by construction flat, by the existence of the crystalline order (and U(1) conservation). The topological $p$-form conservation laws remain valid under arbitrary smooth deformations as long the order persist.  Dislocations represent the vortex-like (dual)disorder operators. On the other hand, also the (local) EM U(1) conservation law is tautological to (gauge) symmetry. Nevertheless, the consequences of the finer details of the theory, as related to the constrained dynamics, can lead to higher-form conservation laws \eqref{eq:lattice_higher_form}, as was discussed with the 3+1d Maxwell theory.

In fact, such a lattice U(1) higher-form theory was described in Ref. \cite{GaiottoEtAl15} and a $\mathbb{Z}_n$ generalization in \cite{GaiottoEtAl17}. In our context, we utilize the ambiguity of line operators on homologically non-trivial cycles. Concretely this amouns to a network od defect operators, such that their presence is consistent with allowed translational charges of the theory. In this case the defect opetators implement large gauge transformations on the Wilson-loop operators $\exp[\im \oint E^a]$, surface operators $\exp[i \oint E^a \wedge E^b]$ etc., which actually define the lattice $L$ via intersections and open sets in the embedding. More abstractly, the construction applies to a topological space with (triangulated) $\check{\rm C}$ech (co)holomology of open covers, or a space admitting a CW-complex up to homotopy equivalence. 

Finally, since the higher-form U(1) theory we consider is constructed with the Wilson lines of $\exp[\im \oint E^a]$ with holonomy fluxes, or equivalently monopole operators realizing the $\mathbb{T}^d$ large gauge transformations along transverse lines, one cannot help to wonder what type of theories could be realized by consdering only suitably defined surface operators and the ensuing generalization of the defect operators in Eq. \eqref{eq:torsion_dislocation}. In some sense, such a higher-form extension is trivially realized if we restrict $E^a \wedge E^b \to E^{ab}$  for some 2-form without lower-dimensional resolution in terms of 1-forms. Below, two-form symmetry without a resolution to 1-form symmetries will correspond to quadrupolar HOTIs in three dimensions. The natural question is whether new classes of responses of higher-form theories would be obtained in this way which are different to those with symmetries \eqref{eq:0-form}, \eqref{eq:1-form}, \eqref{eq:2-form}. Next we classify responses, and their dimensional reductions related to topological states with $p$-form symmetries corresponding to each of these.

\section{Higher form topological crystalline response}\label{sec:generalized_response}

We have discussed the conservation laws related to higher form symmetries above. The response to a $p$-form background field is simple if we identify the corresponding higher $p$-form current $d*J^{(p)} = 0$ \cite{GaiottoEtAl15}, compare to \eqref{eq:Maxwell},
\begin{align}
S^{(p)}[B, J] = \int B^{(p)} \wedge \star J^{(p)} \label{eq:p-form_response}
\end{align}
We now explore such a $p$-form current coupling in the case of the translational crystalline order coupled to the local EM U(1) gauge symmetry. An equivalent problem is the coupling of higher $p$-form gauge fields and currents. However, we focus on the global symmetries, especially since we currently do not now any (simple) incarnations of higher-form gauge fields in Nature, except those from dual magnetic gauge fields. In other words, for the crystalline systems we study, we are interested in the response to the EM U(1) gauge fields in the presence of the (approximate) $p$-form currents $*J^{(p)}$ in Eqs. \eqref{eq:0-form}, \eqref{eq:1-form}, \eqref{eq:2-form}. An important things is that the (lattice) metric enters through the dual lattice $p$-form coupling $\star \to *$ Eq. \eqref{eq:lattice_Hodge}.

The heuristic rule for the relevant higher-form local EM U(1) couplings and responses are suitable substitutions $B^{(p)} \to f^{(p)}(\doo A)$ and $f^{(p)}(A, \doo A)$ in \eqref{eq:p-form_response}, the latter being quadratic in $A$. We now discuss these by stating the results, in terms of the dual, higher-form current pairs. The semi-classical expansion identifying the couplings and associated invariants in momentum space is discussed briefly afterwards.

\subsection{Bulk theta term and charge}
In 3+1d, the simplest crystalline responses corresponds to the 3-form symmetry of lattice volume conservation,
\begin{align}
S_{\theta}[A] &= \int \frac{\theta}{8\pi^2} (* E^1 \wedge E^2 \wedge E^3) F \wedge F \nonumber\\
&= \int \frac{\theta}{8\pi^2} F \wedge F,
\end{align}
where $\theta$ is quantized in the presence of time-reversal symmetry to $0,\pi$ (see also below). Although the crystalline 3-form symmetry superficially trivial/tautological in bulk, it is familiar that the theta term implies protected boundary modes with anomalous $T$-symmetry.

The dual 0-form symmetry response is
\begin{align}
S_{\omega}[E,A] &= \int N_\omega(* 1) \wedge A \nonumber \\
&= \int \frac{N_\omega}{(2\pi)^3} E^1 \wedge E^2 \wedge E^3 \wedge A
\end{align}
corresponding to just the global U(1) charge conservation, where the invariant $N_\omega$ counts the occupied bands in the BZ. Naturally to first order in deformations, $\det(E^{\ a}_i) = 1+\doo_i u^i + O(\doo u^2)$.

These dual responses were (briefly) mentioned  in Ref. \cite{NissinenVolovik2019, Vishwanath2019, NissinenHeikkilaVolovik20} in the present context of elasticity tetrads.

\subsection{Polarization and the quantum Hall effect}

Now we describe responses corresponding to 1-form and 2-form symmetries. The electric polarization \cite{KingSmith93, Resta1994, Resta2010, Rhim17, Watanabe18,Sergeev18, Murakami20, BudichArdonne13} is defined along one-dimensional spatial submanifolds and the quantum Hall response on two-dimensional spatial sections. Both couple to the external electromagnetic field gradient $dA=\frac{1}{2}F_{\mu\nu} dx^{\mu}\wedge dx^{\nu}$. 

\subsubsection{$d+1$-d. polarization}

The 1-form symmetric response is obtained by $B^{(2)} \to dA$.
\begin{align}
S_{\rm pol}[E,A] = 
 \frac{P^a \epsilon_{ab\dots c}}{(d-1)!(2\pi)^{d-1}} \int E^b \wedge \cdots \wedge E^c \wedge dA, \label{eq:polarization}
\end{align}
where the dimensionless coefficient $P^a$ is defined on lattice 1-cycles. In other words, we find that the $P^a$ describes polarization in the direction of the omitted $E^a$ from the response, allowing e.g. open boundary conditions where $a$-periodicity is violated. It is quantized (mod $2\pi$) in terms of protecting (mirror, inversion) symmetries, which are realized anomalously on the boundaries, similar to time-reversal invariant 0-form insulators with quantized theta. Under the large gauge transformations of $E^b \to E^b + \frac{2\pi}{\ell^b}$,
\begin{align}
\delta P^a (\epsilon_{ab\dots d} E^b \wedge\dots\wedge E^d)& = \\
 2\pi P^a (\epsilon_{abc\dots d} & n^b E^c \wedge\dots\wedge E^d)  , a\neq b, \nonumber
\end{align}
where $n^b = L^b/\ell^b$ is the number of lattice points, i.e. density of localized particles, to the $b$-direction. This constraint is tantamount to the integer periodicity of polarization density $P^a(\epsilon_{ab\dots c}E^b \wedge \dots \wedge E^c)$ and was also noted in two-dimensions in \cite{DubinkinEtAl20}.

The polarization response was derived using the crystalline 1-form symmetry but we actually see that it implies also emergent 1-form electric gauge symmetry in the sense that we can shift the local U(1) EM field $A\to A+d\Lambda$ where $\Lambda$ is a 1-form. This viewpoint was studied in Ref. \onlinecite{DubinkinEtAl20}. As usual, we see that here it just a restatement of the crystalline 1-form symmetry $d(*J^{(1)}) = 0$, inherited from the 0-form U(1) charge conservation in the insulator. In a  sense, this the generalization of the 1-form electric symmetry $j = 0$ to the insulator with crystalline symmetries.

We note that $S_{\rm pol}[E,A]$ is a pure boundary term, since
\begin{align}
S_{\doo \rm pol}[E,A] &= \frac{\epsilon_{ab\dots c}}{(d-1)!(2\pi)^{d-1}} \int d(P^a E^b \wedge \cdots \wedge E^c \wedge A)  \nonumber\\
&= \frac{\epsilon_{ab\dots c}}{(d-1)!(2\pi)^{d-1}}  \Delta P^a \int_{\perp^a} E^b \wedge \cdots \wedge E^c \wedge A.
\end{align}
where $\Delta P^a$ is the integrated boundary jump between the insulator and vacuum (for simplicity assumed constant along the transverse directions). We see that the boudary $(d-1)$-volume-form symmetry term is generated. This implies protected boundary modes and the possibility of anomalous Luttinger's theorem on the boundary via anomaly inflow from the bulk to the boundary.

\subsubsection{(d+1)d quantum Hall}
There is a dual response to the polarization which is more familiar. This is simply the (weak) crystalline quantum Hall phase. 
The response follows from the crystalline 2-form symmetry and the replacement $B^{(3)} \to A \wedge dA$, the CS 3-form. It is given as
\begin{align}
S_{\rm QH}[E,A] = \frac{N^{a b}}{2!(d-2)!(2\pi)^{d-1}}  \label{eq:QH_response} \\
\times \int \epsilon_{a b c \dots d} E^c \wedge \cdots \wedge E^d \cdots  \wedge A \wedge dA . \nonumber
\end{align} 
The coefficient $N_{ab}$ is defined on lattice 2-cycles. Note that Refs. \onlinecite{NissinenVolovik2019, NissinenHeikkilaVolovik20} used the dual convention $\tilde{N}_a =\frac{1}{2}\epsilon_{abc}N^{bc}$ for the topological charges for the QHE in 3+1d.

\subsubsection{Discussion}
The expressions are both defined without reference to any spacetime metric.  They depend on (the metric of) the elasticity tetrads $E^a = E_\mu^a dx^{\mu}$ and the embedding with the local deformations. In other words, the responses are not universally quantized. It still follows that the responses are topological in the sections defined by the $d$-cycles of appropriate "missing coordinates" and depend only on the ground state properties which determine the elasticity tetrads. 

The bulk polarization is a total derivative and leads to an emergent electric 1-form gauge symmetry. On the other hand, it is well-known that $S_{\rm QHE}$ is not a total derivative and moreover is not gauge invariant in the presence of boundaries. If we perform 0-form gauge transformation, $A\to A+d\lambda$ the result is the (consistent) anomaly from the bulk to the boundary:
\begin{align}
\delta_{\lambda} S_{\rm QH}[E,A] = \int \frac{1}{2!(d-2)!(2\pi)^{d-1}}  \label{eq:elastic_QHE} \nonumber\\
\times d(N^{a b} \epsilon_{a b c \dots d} E^c \wedge \cdots \wedge E^d \cdots  \wedge \lambda dA)
\end{align}
which, as perhaps expected, now describes a possibly anomalous surface polarization response, where $dN^{ab} \neq 0$ in the absence of dislocations, so that gauge invariance holds overall.

\subsection{Multipole insulators and HOTIs}
Continuing in 3+1d, we now seem to have exhausted the possible responses, since we cannot straightforwardly construct a $p\leq 4$ form out of $A$. On the other hand, the topological responses relevant to multipolar subsystem insulators and electric multipole HOTIs were not yet obtained, except for the polarization HOTI arising from $B^{(2)} \to dA$.

The resolution to the missing ingredient to Eq. \eqref{eq:p-form_response} for such systems is obtained from the fact that we expect additional derivatives of EM fields for multipolar responses, as well as from recent results for HOTIs and higher-rank gauge theories related to crystalline systems. The bulk multipoles themselves satisfy cycle-like conditions with surface multipoles when overlapping at a corner, e.g. for quadropole moment $q_{ij}$ and octopole moment $o_{ijk}$, 
\begin{align}
q_{xy} &= p_x + p_y - q_c \nonumber\\
o_{xyz} &= q_{xy} + q_{yz} + q_{xz} - p_x - p_y - p_z + q_c, \nonumber
\end{align}
etc.\cite{Benalcazar17}, where $q_{ij}$ denotes bulk or surface quadrupole, $p_i$ edge polarizations and  $q_c$ is the corner charge. Related to these, there are the higher-form conservation laws of multipole currents \cite{GromovEtAl20}. In this case the $p$-multipole current couple to higher-order conserved $p+1$ currents, as already discussed. Importantly, for multipole electronic HOTIs, we now require that all lower order bulk multipoles must vanish. In this way, the simple and tautological conserved currents from elasticity tetrads can allow for the correct non-trivial multipole transport. This lead to the form $f^{(p)}(\doo^{p-1} A)$ for $p$-multipoles that enter as $df^{(p)}(\doo^{p-1}A)$ in the response \eqref{eq:p-form_response}. 

Let us now discuss this in three spatial dimensions. We can construct the following 3-form $B^{(3)} \to d f_{q}^{(2)}$,
\begin{align}
f^{(2)}_{q,ij} &= \frac{1}{2}(\doo_i A_j+\doo_i A_j) \sim A_{ij} \\
f^{(2)}_{q,0i} &=  \frac{1}{2} \doo_i A_0 \sim A_{i0} \\ 
(d f^{(2)})_{txy} &= \doo_t A_{xy} - \doo_x \doo_y A_0 \\
& =\frac{1}{2} \doo_t \doo_x A_y + \frac{1}{2}\doo_t \doo_y A_x -\doo_x\doo_y A_0 \\
&= \frac{1}{2}(\doo_x \mathcal{E}_y+\doo_y\mathcal{E}_x). 
\end{align}
The last line follows by EM U(1) gauge invariance (up to finite lattice rotations; we assume simple cubic). The spatial components are obtained from symmetric form $A_{ij}$, as appropriate for multipole moments. From the last identity we see that for multipole response, we couple only to spatial derivates of $A_0$ like for polarization, and the time derivative of the relevant multipole $A_{ij}$.

Plugging this in \eqref{eq:p-form_response}, we obtain a quadrupolar HOTI from crystalline 2-form symmetry
\begin{align}
\frac{\theta\epsilon_{abc}}{2\pi} \int df^{(2)}_{q^{ab}} \wedge E^c .
\end{align}
As for 1-form polarization response, the response is a total derivative. Emergent 2-form gauge symmetry results from $d f^{(2)} \to d f^{(2)} + d\Lambda^{(2)}$ in combination with the conserved current $d*J^{(2)} = d E^z = 0$. The boundary response is, choosing a simple cubic lattice and specialing to the $xy$-plane for simplicity,
\begin{align}
S_{\doo q_{xy}} =  \int \frac{\theta_{xyz}}{2\pi} \left( f^{(2)}_{q_{xy}}(\doo A) \wedge E^z\vert_{y=\rm const.} \right.\\
\left.+  f^{(2)}_{q_{xy}}(\doo A) \wedge E^z\vert_{x=\rm const.} \right)
\end{align}
which is the appropriate for anomalous boundary polarization response, say, along: 
\begin{align}
S_{\doo q_{xy}} &= \frac{\theta_{xy}}{4\pi} \int_{y=\rm const.} \mathcal{E} \wedge E^z\\
&= \frac{\theta_{xy}}{2} \int_{y=\rm const.} dx dt dz \mathcal{E}_x
\end{align}
which naturally leads to polarization $P^a = \theta_{xy}/(2\pi) = 1/2$, in our units, and can been shown to harbor corner charges of $1/4$ if $\theta_{xy}$ is quantized \cite{YouEtAl19}.

In 3+1d we also have the octopole insulator which corresponds to 3-form symmetry without time reversal but instead crystalline symmetries, as distinct from the bulk theta term. In this case $B^{(3)} \to f_{o}^{(3)}(\doo^2A)$ and the response are given as
\begin{align}
S_{o} &= \int \frac{\theta_{abc}}{2\pi} df^{(3)}_{o^{abc}}(\doo^2 A) \\
f^{(3)}_{q,ijk} &= \frac{1}{6}(\doo_i \doo_j A_k+\textrm{symm.}) \sim A_{ijk}, \\
f^{(3)}_{q,0ij} &=  \frac{1}{2}(\doo_i \doo_j A_0 + \textrm{symm.} ) \sim A_{0ij} .
\end{align}
This response was analyzed in detail already in \cite{YouEtAl19}. In this case, since the action is a total derivative and there are no tetrads entering, there seems to be \emph{no} quantization in three-dimensions nor emergent higher-form gauge symmetry with a multipole Chern-Simons. Of course, one could consider the 3+1d octopole insulator to descend from 4+1d pentapole insulator with a multipole Chern-Simons description \cite{YouEtAl19}. 

\subsection{Remarks on higher dimensions}
Naturally the above construction is generalized to arbitrary dimensions by identifying the possible higher-form lattice currents and substituting lower-dimensional topological states in terms of $B^{(p)}\to f^{(p)}(A,\doo A, \dots,\doo^{p-1} A)$. For example, subsystem theta term becomes possible in 4+1d. The responses in  3+1d we considered in detail were just the ones we obtained from gluing the usual theta (or Chern class) and Chern-Simons terms up to that dimension along lattice cycles. For multipoles, we utilized the spatially ``multipolized" higher-derivative forms obtained in multipolar Chern-Simons terms and gauge theories.  Such a construction is directly physically relevant also when considering multipole pumping, essentially by relaxing periodic lattice directions in the currents and momentum space to adiabatic parameters, possibly with non-trivial topology. Natural questions concern the detailed form and realizations of the anomaly inflow, as well as the possibility of realizing states that have non-anomalous response only when considered in combination with higher-dimensional theories (such as e.g. gapless chiral fermions are realized on boundaries of 0-form topological states).

Related to the above, the multipole responses can be generalized to multipole Chern-Simons theories \cite{YouEtAl19}. Here the same construction follows by including additional elasticity tetrads and replacing them with the proper components of (higher-form) gauge field. Such a replacement should potentially arise when integrating out the deformations (quadratically). It would be interesting to consider such multipole Chern-Simons theories in full generality and their dependence on the (background) lattice symmetries and geometry.  On the other hand, we also constructed the \emph{odd} spatial-dimensional axion extension of multipole CS, simply by gluing CS-form with the elasticity tetrads.

\section{The semi-classical expansion and momentum space invariants}\label{sec:momentum_space}

We shall now (briefly) discuss the quantization and momentum space invariants in topological crystalline responses in 3+1d. In this case, the relevant crystalline topological responses are bulk theta-term, polarization, quantum Hall and higher-order multipole responses. For generality, we still write formulas in $d+1$ spacetime dimensions. 

For the multipole responses, the responses are not in general quadratic in the fields, neither elasticity tetrads and the EM gauge field $A$, and the response formulas should be strictly understood to apply for semi-classical, infinitesimal background fields.  We note that even in this limit, the lattice fields $E^a$ are quantized in terms of large gauge transformations.

Accordingly, the invariants computed in the semi-classical expansion in momentum space for weakly varying coordinate deformations and gauge fields is appropriate. This is build on the assumptions are gauge invariance and the validity gradient expansion. Naturally, this expansion can be utilised only in the continuum limit of the elasticity tetrads but we expect that the results are robust to finite lattices and space group symmetries.

Following e.g. Refs. \cite{Volovik2003, QiHughesZhang08, Volovik1988, VayrynenVolovik11}, we imagine that the fermionic degrees of freedom $\Psi$ have been integrated out and focus on the (low-energy) ground state effective action, defined as
\begin{align}
S_{\rm eff}[E, A] = -\im \log \int \mathcal{D}\Psi~ e^{\im S[E,A,\Psi]}
\end{align}
Assuming that the fermions interactions have been decoupled in terms of mean-field Hubbard-Stratonovich fields, the effective action is expanded around saddle-point solutions
\begin{align}
S_{\rm eff}[E,A] = \im \textrm{Tr} \log G[E,A] = -\im \textrm{Tr}  \int^{1}_{0} du~ G \doo_{u} G^{-1} 
\end{align}
where $G[E,A]$ is the Green's function and $E, A$ are background fields. We imagine that the electromagnetic field $A$ is turned on adiabatically as a function of the parameter or extra coordinate $u$ as $A_{u=0} = 0$ and $A_{u=1} = A$. By gauge invariance, $A_{\mu}$ minimally couples to $p_{\mu}$. Similarly, we can define in terms of the adiabatic coordinate $u$,
\begin{align}
E^a_{u=1} &= (E_{\mu}^{(0)a} -E^a_{\mu}(x)) dx^{\mu}, \\
E^a_{u=0} &= E^{(0)a}_{\mu}dx^{\mu} =\frac{2\pi}{\ell^a}\delta^a_{\mu}dx^{\mu}.
\end{align}
where $E^a = \frac{2\pi}{\ell^a} \delta^a_{\mu} dx^{\mu}$, with the convention that $E^0 = \frac{1}{T}\delta^0_t \to 0$ in the case without any periodicity $T$ in the time direction. For the elastiticity tetrads in the response, we simply assume ``geometric gauge invariance". Then the lattice momenta $p^a$ and spacetime momenta $p^{\mu}$ are related by the components, note that former is dimensionless, and $a=1,\dots, d$,
\begin{align}
p^a = E^{a}_{\mu}(x)\hat{p}^{\mu}, \quad p_{a} = G_{ab}\hat{p}^b = E^{\mu}_a(x) \hat{p}_{\mu}.
\end{align}
As a result, the background fields enter as (for a simple, non-interacting system)
\begin{align}
G^{-1}[E,A] &= G^{-1}[p_{a} - A_{a}] = G^{-1}[E_a^{\ \mu}\hat{p}_\mu -A_a]\\
&= \im\omega + A_0 - H[E^i_{\ a}(x)\hat{p}_i-A_a(x)].
\end{align}
We evaluate spacetime derivatives as 
\begin{align}
\doo_{x^{\mu}} G^{-1}[A] & \nonumber\\
= \doo_{k_\nu}& G^{-1} \vert_{u=0} \doo_{x^{\mu}}A_{\nu} +  \doo_{E^i_{\ a}} G^{-1} \vert_{u=0} \doo_{x^{\mu}}E^i_{\ a} \nonumber\\
=  \doo_{k_\nu}& G^{-1} \vert_{u=0} \doo_{x^{\mu}}A_{\nu} + E^{(0)a}_j \hat{p}_i\doo_{p_j} G^{-1}\vert_{u=0} \doo_{x^{\mu}}E^i_{\ a}  \nonumber
\end{align}
where we used $\doo_{A_\mu} G^{-1} = \doo_{k_\mu}G^{-1}\vert_{u=0}$ and $\doo_{E^\lambda_{ a}} G^{-1} =\doo_{p_\mu} G^{-1}\vert_{u=0}E_{\mu}^{(0)a} \hat{p}_\lambda$. We also neglected the second order couplings $A_{a}(x) \approx \hat{E}^{(0)\mu}_a A_\mu = \delta_a^{\mu}A_{\mu}$ in $G^{-1}[E,A]$.

To obtain the sought-for responses, we expand the Green's function $G[E,A]$ in the semi-classical gradient expansion, where $A$ and $E$ are both slowly varying background fields in a double expansion. 

\subsection{Bulk theta term and charge}
These dual invariants are simple and known. The response first order in $A$ zeroth order term in gradients $\doo A$ is
\begin{align}
S_{\rm eff}^{(1,0)}[E,A] = \int^1_0 du \int_{\rm BZ} \frac{d^d\mathbf{p} d\omega}{(2\pi)^{d+1}} \int d^d\mathbf{x} dt  \\
\times \textrm{tr} [G \doo_{k_{\mu}} G^{-1}]_{A=0} \doo_{u}A_{\mu}
\end{align}
Since we assume no periodicity in time, only the integral over lattice spatial directions is well defined. This is just the lattice volume, i.e. the total charge density coupling to $A$,
\begin{align}
S^{(1,0)}[A,E] = \frac{N_1}{d!(2\pi)^d}\int~ \epsilon_{ab\dots c}E^a \wedge E^b \wedge \cdots \wedge E^c \wedge A 
\end{align}
where the invariant corresponding to the 0-form bulk lattice charge is just
\begin{align}
N_\omega = N_\omega(\mathbf{p}) = \frac{1}{2\pi \im} \int_{-\infty}^{\infty} d\omega~ \mathrm{tr} G(p_{\mu}) \doo_{\omega} G^{-1}(p_{\mu}).  \label{eq:filled_bands}
\end{align}
counts the number of occupied states and can only change at BZ momenta $\mathbf{p}$ where the gap closes \cite{Oshikawa2000, Volovik2003, NissinenVolovik2019, Vishwanath2019}. In the actual system, the imaginary frequency is cutoff from above by the charge gap, as well as from below. As discussed, the same response arises on the boundary of the system with non-trivial electric polarization \cite{Vishwanath2019} and is related to the boundary Luttinger anomaly \cite{Vishwanath2019, NissinenVolovik2019}.

For the 3-form crystalline symmetry, the second order in $O(\doo A^2)$ term in the expansion, $S^{(0,2)}_{\rm eff}[A]$ with no elasticity tetrads, producing the bulk theta term with \cite{QiHughesZhang08}
\begin{align}
\theta &= \frac{1}{48\pi^2}\int_0^{2\pi} du \int_{\rm BZ} d\omega d^3 p \epsilon^{u\mu\nu\lambda\rho} \textrm{Tr} \big[(G\partial_u G^{-1}) \\ 
&\times (G\partial_{\mu}G^{-1})(G\partial_{\nu}G^{-1})((G\partial_{\lambda}G^{-1})(G\partial_{\rho}G^{-1})\big] \nonumber
\end{align}
this invariant can be reduced to a pure momentum space winding number, through its relation to the 2nd Chern class of the BZ,
\begin{align}
\theta = \frac{1}{24\pi^2} \int_{\rm BZ} \textrm{Tr}[(GdG^{-1})^3],
\end{align}
making the dual nature to \eqref{eq:filled_bands} more explicit. We also note the relation of these invariants \cite{Volovik2003, Gurarie11}

\subsection{Quantum Hall response}
The semi-classical expansion for the quantum Hall response in terms of elasticity tetrads was discussed in \cite{NissinenVolovik2019}. This term is \cite{Matsuyama1987, Ishikawa1986, Halperin1987, KaplanEtAl93}
\begin{align}
S_{\rm eff}^{(1, 1)}[E,A] = -\frac{1}{4} \int d^{d}\mathbf{x} dt  \int_{\rm BZ} \frac{d^d\mathbf{p} d\omega}{(2\pi)^4} \times  \nonumber \\
 \textrm{tr}[ (G \doo_{k_{\nu}}G^{-1}G \doo_{k_{\mu}}G^{-1} - G \doo_{k_{\mu}}G^{-1}G \doo_{k_{\nu}}G^{-1})\\ 
 G \doo_{p_{\lambda}} G^{-1}]_{u=0} \doo_{\mu}A_{\nu} A_{\lambda}. \nonumber
\end{align}
which is first order in gradients $\doo A$ and second order in $A$. Again the integral splits in momentum-space and the result is Eq. \eqref{eq:QH_response} with the familiar invarant given as
\begin{align}
N_a(k^a) = \frac{1}{8\pi^2}\epsilon_{ijk} \int_{-\infty}^{\infty} d\omega\int_{\rm {BZ}} dS_a^i
\nonumber
\\
{\rm Tr} [(G\partial_{\omega} G^{-1}) (G\partial_{p_j} G^{-1}) ( G\partial_{p_k} G^{-1})]\,.
\end{align}
This invariant is quantized to integers as an element of $\pi_3(\tilde{\mathbb{T}}^3,U(N))$, where $N$ is the number of bands, where $\tilde{\mathbb{T}}^3$ is the doubly pinched 3-torus formed by identifying $\omega\to \pm \infty$ in the sections of 2D BZ transverse to $a$. Under these constraints, also $N_a(k^a)$ is quantized and can change as a function of $k^a$ only when the gap closes in the BZ.

\subsection{Polarization and quadrupolarization}

\subsubsection{Polarization}
We now give an elementary argument for the quantized momentum space polarization invariant inspired by the results \cite{Volovik2003, Gurarie11, EssinGurarie11, VayrynenVolovik11}. Consider a system which is half-open to the $x$-direction with a boundary region at $x=\pm L_x$ with bulk polarization. In the three-dimensional polarization response
\begin{align}
S_{\rm pol} = \frac{1}{2 (4\pi^2)} \int d^3x P^a \epsilon_{abc}E^b \wedge E^c \wedge dA,
\end{align}
we focus on the polarization along $P^1 = P^x$ with $E^1_i = \delta_{ix}$. It is given by
\begin{align}
S_{\rm pol} = \frac{1}{4\pi^2 } \int dydz \int dxdt P^1 E^2_yE^3_z  \epsilon^{\mu\nu yz}\doo_{\mu}A_{\nu}\\
= n_yn_z \int_{-L_x}^{L_x} dx dt (\doo_x P^1) A_0
\end{align}
where for simplicity $A_0$ is constant along $x, y,z$ and the $x$-integral is over the boundary "soliton", where the bulk polarization changes to its vacuum value, and $\int E^2 = n_y, \int E^3 = n_z$. For such a 1d theta term, in the paper \cite{VayrynenVolovik11} the following invariant was given (BZ' denotes the suitably restricted Brillouin zone depending on the symmetries of the Wigner transformed open system)
\begin{align}
N^{3}_{\rm soliton} = \frac{1}{4\pi^2\im} \int_{-\infty}^{\infty} d\omega &  \int_{-L_x}^{L_x} dx \int_{\rm BZ'} dk_x \\ 
\times \tr \big(G \doo_{[x} G^{-1} & G \doo_\omega G^{-1} G \doo_{k_x]} G^{-1} \big) \nonumber \\
=P^1\vert_{L_x} - P^1\vert_{-L_x} & = \frac{1}{2\pi} \int_{\rm BZ'} dk^x~ \tr [G\doo_{k_x}G^{-1}] \nonumber
\end{align}
where, on the last line, we have used the dimensional reduction formula \cite{Gurarie11,EssinGurarie11} at some large imaginary frequency $\omega$. Extending this result to the open-system with two macroscopic boundaries at $x=-L, L$, where $L\to \infty$, and comparing the opposite jumps at both ends, we obtain the result (assuming $P^a = 0$ mod integers, outside the system)
\begin{align}
P^{a} = \frac{1}{2\pi} \int_{\rm BZ} dk^a~ \tr \left(\Sigma G\doo_{k_a} G^{-1}\right). \label{eq:polarization_invariant}
\end{align}
where $\Sigma^2=1$ is an operator that (anti)commutes with the Hamiltonian for e.g. chiral crystalline systems \cite{SchnyderEtAl10}. Remarkably the invariant is well-defined in the \emph{bulk} BZ mod integers and non-trivial if the bulk crystalline symmetry implies $P^a \to -P^a$. In the recent paper \cite{NissinenHeikkilaVolovik20}, this was invariant and boundary response was explicitly studied in $PT$-invariant chiral inversion-time symmetric insulator. In particular, when evaluated for the effective Wigner-transformed Hamiltonian $\mathcal{H}(k_a) = G^{-1}(\omega = 0,k_a)$, which is clearly not well-defined at the edge with gapless flat band excitations. This response is discussed in terms of the semi-classical expansion in the Appendix.

\subsubsection{Quadrupole response}
Let us now discuss the quadrupole response. In contrast to the QH or polarization responses, in this case the response is not obtained from combinations of topological terms $A, dA$ but from the ``multipolelized" electric field form $f^{(2)}(\doo A)$. This term appears in the semi-classical expansion but the resulting momentum space invariant look non-topological, although it will be protected by the $C_4T$ symmetry. This is discussed in Appendix. We also note that since the boundary polarization should be quantized, we expect the bulk invariant to be quantized \cite{YouEtAl19}. 

\section{Conclusions}

We have shown how the topological response of crystalline "embedded" topological insulators and HOTIs coupled to conserved global U(1) symmetries arises in a simple and consistent field theory expansion incorporating the (almost) conserved higher form symmetries related to the gapped insulating lattice. Essentially, the refinement of different insulators was achieved by coupling them to elastic deformations and then considering the possible higher-order \emph{global} symmetries. 

These can appear in the response if the insulator retains the conservation of the charged constituents along some lattice directions. The periodic lattice directions are then embedded to the underlying spacetime using tautological gauge fields, the elasticity tetrads. More formally we can consider lattice theories with networks of monopole operators in Cech (co)homology.

The topological response is particularly clear for the polarization and the quantum Hall invariants, which both descend from lower dimensional topological classes, the 1d theta term and CS term. The latter appears if there is a two-form conservation law along the QH planes, meaning that layers are separated, while the former can appear there are lattice directions along which the charge and polarizations are conserved. The almost conserved symmetry is defined with respected to the charge gap and therefore is not fundamental but emerges at low energy in the topological insulator. The corresponding topological multipoles in the response are similarly well defined if the almost conserved symmetry in the corresponding hyperplanes and the appropriate multipole charge can still move.

In short, the framework is an extension of the topological field theory, in the sense that topological terms are induced with quantized momentum space invariants multiplying by topological terms with mixed field content. The responses are not quantized overall, as can be deduced by simple dimensional analysis, and change under deformations. The corresonding lower-dimensional momentum space invariants should be quantized on the other hand. This requires more work on generalizing the results in \cite{EssinGurarie11} to the higher-order symmetric case. Our framework naturally explains how higher-form \emph{gauge} symmetries can arise in crystals and is equivalent to the reduction of higher rank gauge theories to global U(1) symmetry \cite{DubinkinEtAl20}.  This requires the crystalline symmetries and connects the higher-order gauge symmetry to charge via the almost conserved global symmetries encoded by the elasticity tetrads. The lattice defects carry topological charges and have restricted mobility, whose connection to fractonic phases with higher rank gauge fields has been described recently. In general, the dimensional hierarchy of topological response, subsystem conservation laws and anomalies is realized in a transparent way. 

In the classification of topological crystalline states with symmetry $G$ and ground states $\Omega$, the cohomology $H^{d}(BG,\Omega)$ appears \cite{ThorngrenElse18}. We have focused only on the finite lattice translation symmetries $\mathbb{Z}^d$ in $d+1$-dimensions. The classifying space $BG = EG/G$ is extremely simple and follows as $B\mathbb{Z}^d = \mathbb{T}^d$, when we choose $EG = \mathbb{R}^d$ and fix the origin of the lattice at $X^a=0$. The elasticity tetrads are flat gauge fields on $\mathbb{T}^d$, i.e. elements of $H^{d}(\mathbb{Z},\mathbb{T})$. Here we have outlined the higher-order responses of HOTIs and topological insulators coupled to a conserved U(1) charge. This should be developed to include the space group symmetries and discrete lattices in any real crystalline material. Until now the complete topological classification of crystalline matter has been hindered by the lack of generalization of $K$-theory to include finite space group symmetries. It is possible that the higher-form symmetries and elasticity tetrads could lead to a simpler description of associated vector bundles with appropriate higher rank gauge fields, since the higher-form response is stabilized by the additional crystalline symmetries. The connection to LSM-type theorems should be explored \cite{Oshikawa00, ThorngrenElse18, DubinkinEtAl20}. 

\emph{Note added}: While the current manuscript was prepared, the preprint \cite{DubinkinEtAl20} appeared which discusses HOTIs from the perspective of higher-order \emph{gauge} symmetries. Our results are similar where they overlap.

{\bf Acknowledgements}. I thank T.T. Heikkil\"a and especially G.E. Volovik for discussions and related earlier collaborations. This work has been supported by the European Research Council (ERC) under the European Union's Horizon 2020 research and innovation programme (Grant Agreement No. 694248).

\appendix

\section{Polarization and quadrupole in the semi-classical expansion}
Now we consider the $\doo_u E^a_{\lambda}$ terms in the semiclassical expansion, which heuristically are momentum insertions. These are terms from
\begin{widetext}
\begin{align}
S^{'}_{\rm eff} &=  -\im \Tr \int^{1}_{0} du~ G\doo_{E^\lambda_a}G^{-1}\vert_{u=0} \doo_{u}E^{\lambda}_a = -\im \Tr \int^{1}_{0} du~G\doo_{p_k} G^{-1}\vert_{u=0} E_{k}^{(0)a} \hat{p}_\lambda \doo_{u}E^{\lambda}_a
\end{align}
Now we expand the semiclassical $G$ in the expansion, see e.g. \cite{Gurarie11}. The first, $\doo A$ term is just, with $\doo_{E^\lambda_{ a}} G^{-1} = \delta_{k\lambda}p^a \doo_{p_k} G^{-1}\vert_{u=0}$,
 \begin{align}
S^{(1')} &=  -\im \Tr \int^{1}_{0} du~ G\doo_{E^\lambda_a}G^{-1}\vert_{u=0} \doo_{u}E^{\lambda}_a \\
 &=  -\im \Tr \int^{1}_{0} du~ G\doo_{p_j}G^{-1}\vert_{u=0} E^{(0)a}_j \hat{p}_\lambda \doo_{u}E^{\lambda}_a \\
 &= \int \frac{dtd^d\vek{x}d\omega d^d\vek{p}}{(2\pi)^{d+1}} 
\tr\left[\big(G\doo_{x^\mu} G^{-1}G \doo_{p_{\mu}} G^{-1} 
- G \doo_{p_{\mu}} G^{-1} G \doo_{x^{\mu}} G^{-1}\big)G\doo_{p_j}G^{-1}\right]_{u=0}\hat{p}_j \\
&=\int \frac{dtd^d\vek{x}d\omega d^d\vek{p}}{(2\pi)^{d+1}} 
\tr\left[\big(G\doo_{p_\nu} G^{-1} G \doo_{p_{\mu}} G^{-1} 
- G \doo_{p_{\mu}} G^{-1}G \doo_{p_{\nu}} G^{-1}\big) G\doo_{p_j}G^{-1}\right]_{u=0}\hat{p}_j \doo_{\mu}A_{\nu}.
 \end{align}
The integral is the flux of the 3D invariant $\sim \tr [GdG^{-1}]^3$ over suitable sections of the BZ and can be reduced by the results of \cite{EssinGurarie11} to the one-dimensional invariant in Eq. \eqref{eq:polarization_invariant}, see also \cite{NissinenHeikkilaVolovik20}. 
 
From the semiclassical expansion, second order terms $\doo^2 A$ are from $\doo_\mu \doo_\nu G^{-1} = \doo_{k_{\lambda}}\doo_{k_\rho}G^{-1} \doo_\mu A_{\lambda} \doo_{\nu}A_{\rho} + \doo_{k_{\lambda}}G^{-1}\doo_{\mu}\doo_{\nu}A_{\lambda}$ and for us only the second insertion is important. We get
 \begin{align}
S^{(2')} &= \im \int^{1}_0 du \Tr G\doo_{p_j}G^{-1}\vert_{u=0}  E^{(0)a}_j \hat{p}_\lambda \doo_{u}E^{\lambda}_a \\
&= \frac{\im}{4} \int \frac{dtd^{d}\vek{x}d\omega d^{d}\vek{p}}{(2\pi)^{d+2}}~\tr \bigg[\left(-G\doo_{p_j}G^{-1}G\doo_{p_\mu}G^{-1} G\doo_{p_{\lambda}}G^{-1}G\doo_{p_{\nu}} G^{-1} +G\doo_{p_j}G^{-1}G\doo_{p_\mu}G^{-1} G\doo_{p_{\nu}}G^{-1}G\doo_{p_{\lambda}} G^{-1} 
 \right. \\
& \left. \phantom{GGGGGGGG}+ \doo_{p_{j}}G^{-1} G\doo_{p_{\lambda}}G^{-1} \doo_{p_{\mu}}\doo_{p_{\nu}}G - G\doo_{p_{j}}G^{-1} G\doo_{p_{\mu}} \doo_{p_{\nu}} G^{-1}G\doo_{p_{\lambda}}G^{-1} \right)\bigg]_{u=0}\hat{p}_{j} \doo_{\mu}\doo_{\nu}A_{\lambda}
\end{align}
Terms of the form $-\doo_i \doo_j A_0$ and $\doo_{t} \doo_i A_j$ are produced and the prefactor looks appropriate for a rank-2 multipole but looks superficially non-topological. Moreover, for the $C_4T$-invariant model \cite{Schindler18, YouEtAl19}, 
\begin{align}
C_4&: (t,x,y,z) \to (t,y,-x,z), \quad (A_0,A_x,A_y,A_z) \to (A_0,A_y,-A_x,A_z) \\
T&: t\to -t, \quad (A_0,A_i) \to (A_0, -A_i).
\end{align}
We see that the symmetry fixes $\doo_{\mu}\doo_{\nu} A_{\lambda} = \doo_xE_y +\doo_y E_x$ and corresponding indices on the momentum space factors. Naturally, in order to have a non-trivial response, the momentum space integral has to respect the same symmetries. It is possible that the invariant can be reduced in the Brillouin zone to a 3D invariant, see below Eq. \eqref{eq:theta_q}. We should also note that similar term quadratic in $A_{\mu}$ is produced, with the $\hat{p}_j \to A_j$ in the expansion.

In the paper \cite{SchindlerEtAl18}, this response was related to a non-constant theta term, now quantized with respect the $C_4T$ symmetry $\theta(x,y,z,t) \to \theta(y,-x,z,t)$, similar to the TRI invariant $\mathbb{Z}_2$ insulator with theta term quantized in the bulk and anomalous at the boundary. We now want to connect this response to our results. 
For a spacetime dependent $\theta(x)$, the action making is sense is really the one corresponding to the 3+1d QH,
\begin{align}
S'_{q_{xy}}[A,\theta] = \frac{1}{16\pi^2} \int d^4x~ \epsilon^{\alpha\beta\gamma\delta}A_{\alpha}F_{\beta\gamma}\doo_{\delta}\theta .
\end{align}
with the identification $\doo_\delta \theta dx^{\delta} = \frac{1}{2}\epsilon^{xy}\omega_{xy}$, an ``axionic" disclination, instead of the topological invariant $F\wedge F$. Due to the $C_4T$-symmetry, $\theta$ winds by $2\pi$-jumps in the four quadrants, which can be pushed by deformation to the as $\pm2\pi$ vortices corners. The singularities $(\doo_\alpha \doo_{\beta}-\doo_{\beta}\doo_\alpha)\theta \neq 0 = (d\omega)_{\alpha\beta}$, localized on the corners in the $xy$-plane for $q_{xy}$-quadrupole. This gives QH-like currents, polarizations and magnetizations, as in 2+1d, 
\begin{align}
j^\alpha &= \frac{\delta S'_{q_{xy}}}{\delta A_{\alpha}} = \frac{1}{8\pi^2}\epsilon^{\alpha\beta\gamma\delta} F_{\beta\gamma}\doo_\delta \theta,\\ 
P^{i} &= \frac{\delta{S'_{q_{xy}}}}{\delta \mathcal{E}_i} = \frac{1}{8\pi^2}\epsilon^{0ijk} A_{j}\doo_{k}\theta\\
M^{i} &= \frac{\delta{S'_{q_{xy}}}}{\delta \mathcal{B}_i} = \frac{1}{8\pi^2}A_{0}\doo_{i}\theta.
\end{align}
In Ref. \cite{VayrynenVolovik11} the following solitonic formula, e.g. at the boundary of the system, was considered,
\begin{align}
\doo_{\mu}\theta = \frac{1}{2\pi^2\im} \int_{\rm BZ} & d^3kd\omega~ \epsilon^{0lmn}\tr \big(G\doo_\mu G^{-1} \\
\times G \doo_\omega G^{-1} G\doo_{k_l} & G^{-1} G\doo_{k_m}G^{-1} G\doo_{k_n}G^{-1} \big) \\
=\frac{1}{10\pi^2\im} & \int_{\rm BZ}  d\omega d^3k \doo_{\mu}n_{5}^{\mu},
\end{align}
where $n^{\mu}$ is defined by the above formula. This should now be replaced with the 7-form winding formula, which also depends on the coordinates $x, y,z$ in the presence of boundaries (e.g. along $z$):
\begin{align}
\Delta \theta= \int dz \doo_{\mu}\theta 
=\frac{1}{(2\pi)^4\im} \int dz \int_{\rm BZ} & d\omega d^3k \doo_{\mu}n_{5}^{\mu} \nonumber,
\end{align}
Applying the same logic as before for the polarization for the $z$-boundaries of an open system, reducing in the $z,\omega$-dimensions, the formula becomes
\begin{align}
\theta_{\rm q-pole} &= \int_{\rm BZ} d^3 k \epsilon^{lmn}\tr\big( G\doo_\mu G^{-1} \label{eq:theta_q}\\
\times G \doo_\omega G^{-1} G\doo_{k_l} & G^{-1} G\doo_{k_m}G^{-1} G\doo_{k_n}G^{-1} \big) \nonumber
\end{align}
where the Green's function depends on the coordinates $(\omega, \vek{k}; x,y,z)$.
 \end{widetext}



 \end{document}